\documentclass[a4paper]{article}

\usepackage{INTERSPEECH2016}

\usepackage{graphicx}
\usepackage{amssymb,amsmath,bm}
\usepackage{textcomp}
\usepackage{array}
\usepackage{stfloats}

\ninept

\title{Binary Speaker Embedding}

\makeatletter
\def\name#1{\gdef\@name{#1\\}}
\makeatother \name{{\em Lantian Li, Chao Xing, Dong Wang, Kaimin Yu, Thomas Fang Zheng$^*$\thanks{{This work was supported by the National Natural Science Foundation of China under Grant No. 61371136 and No. 61271389, it was also supported by the National Basic Research Program  (973 Program) of China under Grant No. 2013CB329302. D.W. and T.F.Z. are with Division of Technical Innovation and Development of Tsinghua National Laboratory for Information Science and Technology and Research Institute of Information Technology (RIIT) of Tsinghua University. This paper is also supported by Pachira.}}}}

\address{Center for Speech and Language Technologies, Tsinghua University \\
    {\small \tt \{lilt,xingchao,yukm\}@cslt.riit.tsinghua.edu.cn; wangdong99@mails.tsinghua.edu.cn} \\
    {\small \tt $^*$Corresponding Author:fzheng@tsinghua.edu.cn}}
\begin{document}

\newcommand{\tabincell}[2]
{

}
\maketitle
\begin{abstract}

The popular i-vector model represents speakers as low-dimensional continuous vectors (i-vectors),
and hence it is a way of continuous speaker embedding.
In this paper, we investigate binary speaker embedding, which transforms i-vectors
to binary vectors (codes) by a hash function.  We start from locality sensitive hashing (LSH), a simple binarization approach where binary
codes are derived from a set of random hash functions. A potential problem of LSH is that the randomly sampled hash functions
might be suboptimal. We therefore propose an improved Hamming distance learning approach, where the hash function is learned by a variable-sized
block training that projects each dimension of the original i-vectors to variable-sized binary codes independently.

Our experiments show that binary speaker embedding can deliver competitive or even better results on both speaker verification and identification tasks,
while the memory usage and the computation cost are significantly reduced.

\end{abstract}

\section{Introduction}
\label{sec:intro}

The popular i-vector model for speaker recognition assumes that a speech segment can be represented
as a low-dimensional continuous vector (i-vector) in a subspace that involves both speaker and
channel variances~\cite{ES3,ES4}.
Normally the cosine distance is used as the distance measure in this i-vector space. Various discrimination
or normalization approaches have been proposed to improve the i-vector model, e.g.,
linear discriminant analysis (LDA)~\cite{ES1}, within-class covariance normalization (WCCN)~\cite{ES7},
probabilistic linear discriminant analysis (PLDA)~\cite{ES6}. We prefer LDA because it is simple and effective,
achieving similar performance as the complex PLDA while preserving the simple scoring based on cosine distance,
which is highly
important for large-scale applications. In this paper, whenever we mention the i-vector model
or i-vectors, we mean i-vectors with LDA employed.

The i-vector model can be regarded as a \emph{continuous} speaker embedding, which projects a complex and
high-dimensional structural data (speech signal) to a simple speaker space that is low-dimensional and continuous.
Despite the broad success of this approach, there are some potential problems associated with the continuous embedding.
Firstly, although i-vectors are quite compact representations of speakers (compared to the conventional
Gaussian mixture models (GMMs)), memory usage and computation cost are still demanding for large-scale tasks.
For example, if the dimensionality of an i-vector is $150$ and each dimension is a float ($8$ bytes), representing
one billion people (the population of China) requires $1.2$ TB memory. To search for a people given a reference i-vector,
the computation cost involves one billion cosine distance calculations, which is very demanding. Note that the
computation will become prohibitive
if the model is based on GMMs or if the scoring is based on PLDA, that is why we focus on the LDA-projected i-vector model in this paper.

Another potential problem of the continuous speaker embedding, as we conjecture, is the over sensitivity to non-speaker
variances. We argue that since the vectors are continuous and can be changed by any small variances in the speech signal,
i-vectors tend to be `over representative' for subtle information that are irrelevant to speakers. LDA can partly solve
this problem, but it is the nature of the continuous representation that makes it fragile with corruptions.
This resembles to the fact that analog signals tend to be impacted by transmission errors.

In this paper, we propose to use binary speaker embedding to solve the above problems. More specifically, we transfer
i-vectors to binary vectors (codes) on the principle that the cosine distance in the original i-vector space is
largely preserved in the new binary space measured by the Hamming distance. The binary embedding leads to significant
reduction in storage and computing cost; additionally, since binary vectors are less sensitive to subtle change,
we expect more robustness in conditions with noise or channel mismatch.

We start from the simple binary embedding method based on locality sensitive hashing (LSH)~\cite{gionis1999similarity,charikar2002similarity,andoni2006near},
and then extend to a Hamming distance learning method~\cite{ES8}. Particularly, we propose a variable-sized block training algorithm
that can improve the learning speed and allocate more bits for important dimensions.

One may argue that the binary embedding is a retraction back to the historical one-hot encoding,
and binary codes are less representative than continuous vectors unless a very large dimensionality is used.
However, our experiments showed that this is not the truth: very compact binary vectors can represent tens of thousands
of speakers pretty well, and binary vectors work even better in some circumstances.
These observations indicate that binary embedding is not an odd retraction to the one-hot encoding; it is essentially a simple
speaker information distillation via hashing.

The rest of this paper is organized as follows. Section~\ref{sec:rel} describes the related work;
Section~\ref{sec:lsh} presents the LSH-based binary embedding, and Section~\ref{sec:var} presents the variable-sized block training.
The experiments are presented in Section~\ref{sec:exp}, and Section~\ref{sec:conl} concludes the paper.

\section{Related work}
\label{sec:rel}

Binary embedding has not been fully recognized in the speaker recognition community. The limited research focuses on employing
the advantages of binary codes in robustness and fast computing. For example, \cite{shao2006robust} proposed a time-spectral
binary masking approach to improve robustness of speaker recognition in conditions with high interference.
Besides, \cite{ryan2014lsh} presented a solution for large-scale speaker search and indexing under the i-vector model,
where the search and indexing algorithm is based on LSH.
The work proposed in~\cite{bonastre2011discriminant} is more relevant to our proposal. By their approach, a universal background model (UBM)
is employed to divide the acoustic space into subregions, and each subregion is populated with a set of Gaussian components.
Each acoustic frame is then converted to a binary vector by evaluating the Gaussian components that the frame belongs to, and the
frame-level vectors are finally accumulated to produce the segment-level speaker vector. Better robustness compared
with the conventional GMM-UBM approach was reported by the authors.

\section{Binary speaker embedding with LSH}
\label{sec:lsh}

We present the binary embedding approach for speaker recognition. Basically the continuous i-vectors are projected to binary codes in such a way that
the distance between i-vectors is largely preserved by the binary codes. We consider the cosine distance for i-vectors (which is the most simple and
effective for speaker recognition) and the Hamming distance for binary codes (which is the most popular distance measure for binary codes).

Let $x$ denote a length-normalized i-vector, and the similarity between i-vectors is measured by the cosine distance. Our goal is to project a continuous vector $x$ to
a binary code $h(x)$ of $b$ bits.
The LSH approach~\cite{gionis1999similarity,charikar2002similarity,andoni2006near} seeks for a hash function operating on $x$,
such that more similar i-vectors have more chance to coincide after hashing.

We employ a simple LSH approach proposed in~\cite{charikar2002similarity}. It selects $b$ hash functions $h_r(\cdot)$, each of which simply
rounds the output of the product of $x$ with a random hyperplane defined by a random vector $r$:
\vspace{-1mm}
\begin{equation}
h_{r}(x)=\left\{
\begin{array}{rcl}
1     &      if \ r^{T}x \geq 0 \\
0     &      otherwise \\
\end{array} \right.
\label{eq1}
\end{equation}

\noindent where $r$ is sampled from a zero-mean multivariate Gaussian $N(0; I)$.
It was shown by~\cite{goemans1995improved} that the following LSH requirement is satisfied:
\vspace{-1mm}
\begin{equation}
    P[h(x_i) = h(x_j)] = 1- \frac{1}{\pi} \theta(x_i, x_j)
\label{eq2}
\end{equation}

\noindent where $\theta(x_i, x_j)$ is the angle between $x_i$ and $x_j$ and is closely related to their cosine distance. Intuitively, this
means that similar i-vectors have more chance to
be encoded by the same binary vector than dissimilar ones, which just coincides our goal of preserving similarities of i-vectors with the binary codes.

\section{Binary embedding with variable-sized block training}
\label{sec:var}

A potential problem of the LSH embedding is that $x$ is not necessarily uniformly distributed on the hyper sphere, and so
the uniformly sampled hash functions $\{h_r\}$ might be suboptimal.
A better approach is to derive the hash function by learning from data.
An interesting method of this category is the Hamming distance learning proposed by~\cite{ES8}. This section presents this approach first,
and then proposes a variable-sized block training method that can improve training speed and quality.

\subsection{Hamming distance learning}

The Hamming distance learning approach~\cite{ES8} learns a projection function $f(x;w)$ where $x$ is the input (an i-vector in our case)
and $w$ is the model parameter.
Once the projection function is learned, the binary code for $x$ is obtained simply by $b(x;w)$ = $sign(f(x;w))$.
Choosing different $f$ leads to different learning methods. The simple linear model $f(x;w)$ = $w^{T}x$ is chosen in this
study. Note that if $w$ is randomly sampled from $N(0;I)$ and no training is performed, this approach is equivalent to LSH.

The Hamming distance learning defines a loss function on triplets $(x,x^{+},x^{-})$, where $x$ is an i-vector of a particular speaker, $x^{+}$
is another i-vector of the same speaker derived from a different speech segment, and $x^{-}$ is the i-vector of an imposter.
The goal of Hamming distance learning is to optimize $w$ such that $b(x;w)$ is closer to $b(x^{+};w)$ than $b(x^{-};w)$ in terms of Hamming distance.
Denoting $(h,h^{+},h^{-})$ as the binary codes obtained by applying $b(x,w)$ to the triplet $(x,x^{+},x^{-})$, the loss function of the learning is:

\vspace{-1mm}
\begin{equation}
l(h,h^{+},h^{-}) = [||h-h^{+}||_{H}-||h-h^{-}||_{H} + 1]_{+}
\label{eq3}
\end{equation}

\noindent where $||\cdot||_{H}$ is the Hamming distance, defined as the number of $1's$ in the vector. Adding the loss function and a regularization term,
the training objective function with respect to $w$ is defined as follows:
\vspace{-1mm}
\begin{equation}
    L(w) = \sum_{(x,x^{+},x^{-}) \in D} l(b(x;w),b(x^{+};w),b(x^{-};w)) + \frac{\lambda}{2}||w||^2
\label{eq4}
\end{equation}

\noindent where $D = \{(x_{i},x_{i}^{+},x_{i}^{-})\}_{i=1}^n$ denotes the training samples, and $\lambda$ is a factor to scale the contribution of the regularization term.
Note that this approach has been employed to image retrieval in~\cite{ES8}, though in this paper we use it for speaker recognition.

\subsection{Variable-sized block training}

A particular problem of the Hamming distance learning is the high computation demand if the dimensions of the continuous and/or binary vector are large.
Additionally, the learning algorithm treats each dimension of the input continuous vector equally, which is not optimal for the LDA-projected i-vectors
for which the low dimensions involve more discriminative information. We propose a variable-sized blocking training approach to solve this problem.

Considering that the expected number of bits of the binary codes is $b$, we hope these bits are distributed to the dimensions of the original
i-vectors unequally, subjected to the constraints $\sum_{i=1}^D{T_{i}} = b$ where $D$ is the dimensionality of the original i-vectors, and ${T_{i}}$ is the number of bits allocated to dimension $i$.  $T_i$ is designed to be linearly descended as follows:

\begin{equation}
    T_{i} = \frac{D+1-i}{D}T_{1}
\label{eq5}
\end{equation}

\noindent This leads to $T_{i} = \frac{2b(D+1-i)}{D(D+1)}$, and the ceil value $T_{i} = \lceil\frac{2b(D+1-i)}{D(D+1)}\rceil $ is selected as the number of encoding bits for the $i$-th dimension.

Specifically, the variable-sized block training first defines the number of bits $T_i$, and then the Hamming distance learning is employed to learn the projection matrix $w_i$ for the $i$-th dimension. The learned $w_i$ is used to embed
the $i$-th dimension of the i-vectors to binary codes. Since the learning and embedding for every dimension $i$ is independent, this in fact leads to a block diagonal parameter matrix $w$ (so the block training is named):
\[
w = \begin{pmatrix}
w_1 & 0 & 0 &\cdots &0 \\
0   & w_2&0 &\cdots & 0 \\
\vdots & \vdots & \ddots & \vdots \\
0   & 0  & 0 & \cdots &w_D
\end{pmatrix}.
\]

Note that this block training learns each dimension independently so it is faster than the conventional Hamming distance learning where the
projection matrix $w$ is learned as a whole. Additionally, because more bits are allocated for low dimensions (which
involve more information due to LDA), the resultant binary codes are more representative and discriminative.

\section{Experiments}
\label{sec:exp}

The proposed binary embedding approach was tested on both speaker verification and identification tasks.
We first present the data and configurations used in the experiments, and then report the results on the verification and identification tasks respectively.

\subsection{Data}

\begin{itemize}
    \item \textbf{Development data}:

\begin{itemize}
    \item \noindent Fisher: $7,196$ female speakers with $13,287$ utterances were used to train the i-vector, LDA models.
    The same data were also used to conduct the variable-sized block training.
\end{itemize}
\end{itemize}

\begin{itemize}
    \item \textbf{Evaluation data}:

\begin{itemize}
        \item NIST SRE08: The data of the NIST SRE08 core test in short2 and short3 conditions~\cite{ES9} were used for the speaker verification evaluation. It consists of $1,997$ female enrollment utterances and $3,858$ test utterances. We constructed $59,343$ trials based on the database, including $12,159$ target trials and $47,184$ imposter trials.

        \item WSJ: The WSJ database was used for the speaker identification evaluation. It consists of $282$ female speakers and $37,317$ utterances. For each speaker, $5$ utterances were randomly selected to
            train the speaker models, and the remaining utterances were used for evaluation, including $35,907$ test trials.

\end{itemize}
\end{itemize}

\subsection{Experimental setup}

The acoustic feature involved $19$-dimensional Mel frequency cepstral coefficients (MFCCs) together with the log energy.
The first and second order derivatives were augmented to the static features, resulting in $60$-dimensional feature vectors.
The UBM involved $2,048$ Gaussian components and was trained with about $8,000$ female utterances selected from the Fisher database randomly. The dimensionality of the i-vectors was $400$.
The LDA model was trained with utterances of $7,196$ female speakers, again randomly selected from the Fisher database.
The dimensionality of the LDA projection space was set to $150$.
For the variable-sized block training, utterances in the Fisher database were sampled randomly to build the contrastive triples and were used to train the projection function.

\subsection{Speaker verification task}
\label{sec:sv-task}

The first experiment investigates the performance of binary speaker embedding on the speaker verification task. All the i-vectors have been transformed by
LDA, and the dimensionality is $150$.  The performance is evaluated in terms of equal error rate (EER) under the NIST SRE08 evaluation set, and the results are shown in Table~\ref{tab:svsre08-lsh}
for the LSH approach, and Table~\ref{tab:svsre08-var} for the variable-sized block training. In each table, the performance with binary codes (denoted by `b-vector')
of various sizes
are reported. Note that we didn't report the time cost in this experiment
since the computation is not a serious problem in speaker verification, although binary vectors are certainly faster.

\begin{table}[htp]
        \centering
          \caption{\it EER\% with LSH-based binary embedding.}
          \label{tab:svsre08-lsh}
          \vspace{2mm}
         \begin{tabular}{l|c|c|c|c|c}
            \hline
                           & i-vector    &  \multicolumn{4}{c}{ b-vector}         \\
            \hline
              Bits         &   9600       &   150   &   300   &   600    & 900       \\
            \hline
              C1       &   {\bf 20.80}    &  26.57  &  24.93  &  22.94   & 21.69   \\
              C2       &   {\bf 1.79}     &  5.97   &  4.18   &  2.98    & 2.09  \\
              C3       &   {\bf 20.97}    &  27.95  &  25.57  &  23.03   & 21.72   \\
              C4       &   {\bf 13.21}    &  26.43  &  21.47  &  17.12   & 16.07   \\
              C5       &   {\bf 14.78}    &  24.52  &  21.15  &  18.15   & 17.55   \\
              C6       &   {\bf 9.92}     &  15.80  &  13.80  &  12.25   & 11.25   \\
              C7       &   {\bf 5.58}     &  12.42  &  9.63   &  8.11    & 7.22  \\
              C8       &   {\bf 6.32}     &  12.37  &  9.74   &  8.16    & 7.11  \\
           \hline
              Overall  &   {\bf 20.24}    &  23.21  &  22.05  &  21.02   & 20.61  \\
            \hline
          \end{tabular}
\end{table}

\vspace{-1mm}

\begin{table}[htp]
        \centering
        \caption{\it EER\% with variable-sized block training.}
          \label{tab:svsre08-var}
          \vspace{2mm}
          \begin{tabular}{l|c|c|c|c|c}
            \hline
                           & i-vector    &  \multicolumn{4}{c}{ b-vector}              \\
            \hline
              Bits         &   9600      &   150    &   300    &   600   &  900         \\
            \hline
              C1       &   20.80         &  24.46   &  20.82   &  19.67  & {\bf 19.39}   \\
              C2       &   {\bf 1.79}    &   5.67   &   4.78   &   3.28  &   2.98       \\
              C3       &   20.97         &  25.73   &  21.29   &  19.91  & {\bf 19.80}   \\
              C4       &   {\bf 13.21}   &  23.42   &  16.82   &  16.37  &   15.32       \\
              C5       &   {\bf 14.78}   &  23.32   &  17.55   &  16.23  &   15.99       \\
              C6       &   {\bf 9.92}    &  14.58   &  10.92   &  10.59  &   10.42      \\
              C7       &   {\bf 5.58}    &  10.77   &   7.86   &   7.10  &   6.84      \\
              C8       &   {\bf 6.32}    &  11.32   &   7.89   &   8.16  &   7.63      \\
            \hline
              Overall  &   20.24         &  21.86   &  19.54   &  18.67  & {\bf 18.64}  \\
            \hline
          \end{tabular}
\end{table}

From the results in Table~\ref{tab:svsre08-lsh} and Table~\ref{tab:svsre08-var}, it can be observed that binary vectors can achieve
performance comparable to the conventional i-vectors, in spite of the much smaller number of bits. For example, with the largest binary codes,
the number of bits is only one tenth of that of the original i-vectors. Compared the two binary embedding methods, it is clear
that the variable-sized block training performs better consistently. In condition $1$ and $3$, the binary codes derived
by the variable-sized block training work even better than the i-vectors. Note that the conditions  where the binary codes perform
better than i-vectors are all with microphones, which are different from the condition of the training data (Fisher database that was recorded by telephones). This seems to support our conjecture that binary codes are more robust to speaker-irrelevant variations.

\subsection{Speaker identification task}

The advantage of the binary embedding is more evident on the speaker identification task, where significant
computation is required when computing the
k-nearest candidates of a given speaker vector. We use the WSJ database for evaluation, which contains $282$ female speakers, and $35,907$ target trials.
For each trial $(x,y) \in V$, where $V$ is the speaker correspondence set, $x$ is the enrollment speaker vector and $y$ is the test speaker vector.
In speaker identification, given a test utterance $y$ whose speaker vector is $x$, the task is to search for the k-nearest speaker vectors around $x$. If
a vector $y$ is in the k-nearest candidates and $(x,y)$ is in the speaker correspondence set $V$, then a top-k hit is obtained. We evaluate the performance
of speaker identification by the top-k accuracy, which is defined as the proportion of the top-k hits in all the trials. Note that we use only a naive
k-nearest search which calculates the distance of the test vector to all the speaker vectors and select the k-nearest candidates. In fact,
various methods can be employed to improve efficiency of the search in particular for binary codes, e.g., the PLEB algorithm ~\cite{indyk1998approximate,charikar2002similarity}. We focus on computation cost of the basic algorithm in this paper.

\begin{table}
        \centering
        \vspace{-1mm}
          \caption{\it Top-k accuracy (Acc\%) with binary embedding based on LSH.}
          \label{tab:sidsre08-lsh}
          \vspace{2mm}
          \begin{tabular}{l|c|c|c|c|c}
            \hline
                           & i-vector     &  \multicolumn{3}{c}{ b-vector}         \\
            \hline
              Bits         &   9600       &   150     &   300     &   600     &   900    \\
            \hline
                Top-1      & {\bf 92.87}  &   55.85   &   73.35   &   83.44   &   86.87   \\
            \hline
                Top-3      & {\bf 97.00}  &   72.17   &   85.76   &   91.98   &   94.00  \\
            \hline
                Top-5      & {\bf 98.01}  &   78.56   &   89.58   &   94.39   &   95.86  \\
            \hline
                Top-10     & {\bf 98.94}  &   85.84   &   93.52   &   96.61   &   97.63  \\
            \hline
                Speed up   &  $\times$1   & {\bf$\times$287}  & {\bf$\times$184}  & {\bf$\times$94} & {\bf$\times$66}    \\
            \hline
          \end{tabular}
\end{table}

\begin{table}[htp]
        \centering
        \vspace{-1mm}
          \caption{\it Top-k accuracy (Acc\%) with binary embedding based on variable-sized block training.}
          \label{tab:sidsre08-var}
          \vspace{2mm}
          \begin{tabular}{l|c|c|c|c|c}
            \hline
                           & i-vector    &  \multicolumn{3}{c}{ b-vector}         \\
            \hline
              Bits         &   9600      &   150       &    300        &    600      &  900       \\
            \hline
                Top-1      & {\bf 92.87}  &   55.94    &    75.57      &    79.58    &  79.96   \\
            \hline
                Top-3      & {\bf 97.00}  &   71.18    &    86.67      &    89.53    &  89.87   \\
            \hline
                Top-5      & {\bf 98.01}  &   77.30    &    90.31      &    92.38    &  92.70   \\
            \hline
                Top-10     & {\bf 98.94}  &   84.35    &    94.04      &    95.41    &  95.69         \\
            \hline
              Speed up    &  $\times$1   & {\bf$\times$287}  & {\bf$\times$184}  & {\bf$\times$94} & {\bf$\times$66}    \\
            \hline
          \end{tabular}
\end{table}

The top-k accuracy with the two binary embedding approaches are reported in Table~\ref{tab:sidsre08-lsh} and Table~\ref{tab:sidsre08-var}, respectively.
For comparison, the bits of the vectors and the computation cost (relative to the i-vector system) are also reported.
From these results, we observe that binary vectors can approach performance of the conventional i-vectors with much fewer bits and
much faster computation. Compared to LSH, the variable-sized block training leads to slightly worse performance.
We attribute this result to the fact that the objective function of the variable-sized block training is pair-wised discrimination (true or feigned speakers),
which is not directly related to the metric in speaker identification.
The results we obtained in Table~\ref{tab:sidsre08-lsh} and Table~\ref{tab:sidsre08-var} clearly demonstrate that
the binary embedding performs much faster than the conventional continuous embedding,
and thus is highly suitable for large-scale identification tasks, e.g., national-wide criminal search.

\section{Conclusions}
\label{sec:conl}

This paper investigated the binary embedding approach for speaker recognition.
We studied two binarization approaches, one is based on LSH and
the other is based on Hamming distance learning. Our experiments on both speaker verification and identification tasks show that
binary speaker vectors can deliver competitive results with smaller vectors and less computation compared to the conventional i-vectors.
This is particularly true with the proposed variable-sized block training algorithm, an extension of the
conventional Hamming distance learning method.

Although it has not completely beat the continuous i-vectors, the binary speaker embedding proposed in this paper is still very promising.
Future work will study more powerful methods to learn the hash function, and investigate the methods to learn binary vectors from speech signals directly.

\eightpt
\bibliographystyle{IEEEtran}

\bibliography{binary}

\end{document}